\documentclass[twocolumn,showpacs,preprintnumbers,amsmath,amssymb,superscriptaddress,groupedaddress,prl]{revtex4}

\pdfoutput=1                  
\usepackage{epsfig}                                                            
\usepackage{graphicx}
\usepackage{dcolumn}
\usepackage{color}
\usepackage{bm}
\usepackage{subfigure} 
\usepackage{epsfig}
\usepackage{fontenc} 

\begin{document}

\title{\bf Reply to the comment on "Superspin Glass Mediated Giant Spontaneous Exchange Bias in BiFeO$_3$-Bi$_2$Fe$_4$O$_9$ Nanocomposite"}

\author {Tuhin Maity} \affiliation {Micropower-Nanomagnetics Group, Microsystems Center, Tyndall National Institute, University College Cork, Lee Maltings, Dyke Parade, Cork, Ireland}
\author {Sudipta Goswami} \affiliation {Nanostructured Materials Division, CSIR-Central Glass and Ceramic Research Institute, Kolkata 700032, India}
\author {Dipten Bhattacharya} 
\email{dipten@cgcri.res.in} \affiliation {Nanostructured Materials Division, CSIR-Central Glass and Ceramic Research Institute, Kolkata 700032, India}
\author {Saibal Roy}
\email{saibal.roy@tyndall.ie} \affiliation {Micropower-Nanomagnetics Group, Microsystems Center, Tyndall National Institute, University College Cork, Lee Maltings, Dyke Parade, Cork, Ireland} 

\date{\today}

\begin{abstract}
In this article we reply to the concerns raised by Harres $\textit{et al}$. [Phys. Rev. Lett. (to be published)] about some of the results reported in our original paper [T. Maity $\textit{et al}$. Phys. Rev. Lett. $\textbf{110}$, 107201 (2013)]. We show that the magnetic hysteresis loops are not minor and both path dependency of exchange bias and presence of superspin glass phase in the nanocomposite are indisputable.
\end{abstract} 
\pacs{75.70.Cn, 75.75.-c}
\maketitle

In their Comment, Harres $\textit{et al}$. \cite{Harres} expressed concerns about some of the results reported in our paper \cite{Maity-1}. In this Reply, we take up and address their comments point by point. The first and the most important issue is whether the hysteresis loops are minor. A minor loop is nonsymmetrical and is expected to exhibit significant vertical asymmetry as well \cite{Wang, Klein}. The loops reported in our work do not exhibit any vertical asymmetry. The magnetizations corresponding to the positive and negative maximum fields ($H_m$) are equal (Fig. 1) within the field and temperature limits of our study. Moreover, the magnetization is clearly found to be reversible (magnetizations for forward and reverse branches of the loop are merged around the maximum field applied) in the present case (Fig. 1). Fig. 2 shows the blown up portions of the hysteresis loops around the high field region. It is true that within the field limit applied, complete saturation of the magnetization could not be observed due to the presence of antiferromagnetic BiFeO$_3$ by a larger volume fraction. Incomplete saturation has been observed in a variety of systems where ferromagnetism is weak and associated with other magnetic phases, especially, spin glass. The spin structure evolves here continuously with the field and the exchange bias is found to depend on the applied field \cite{Wang, Kleemann}. Hence, the doubt raised in the comment is not well founded. 

\begin{figure}[!h]
  \begin{center}
    \includegraphics[scale=0.15]{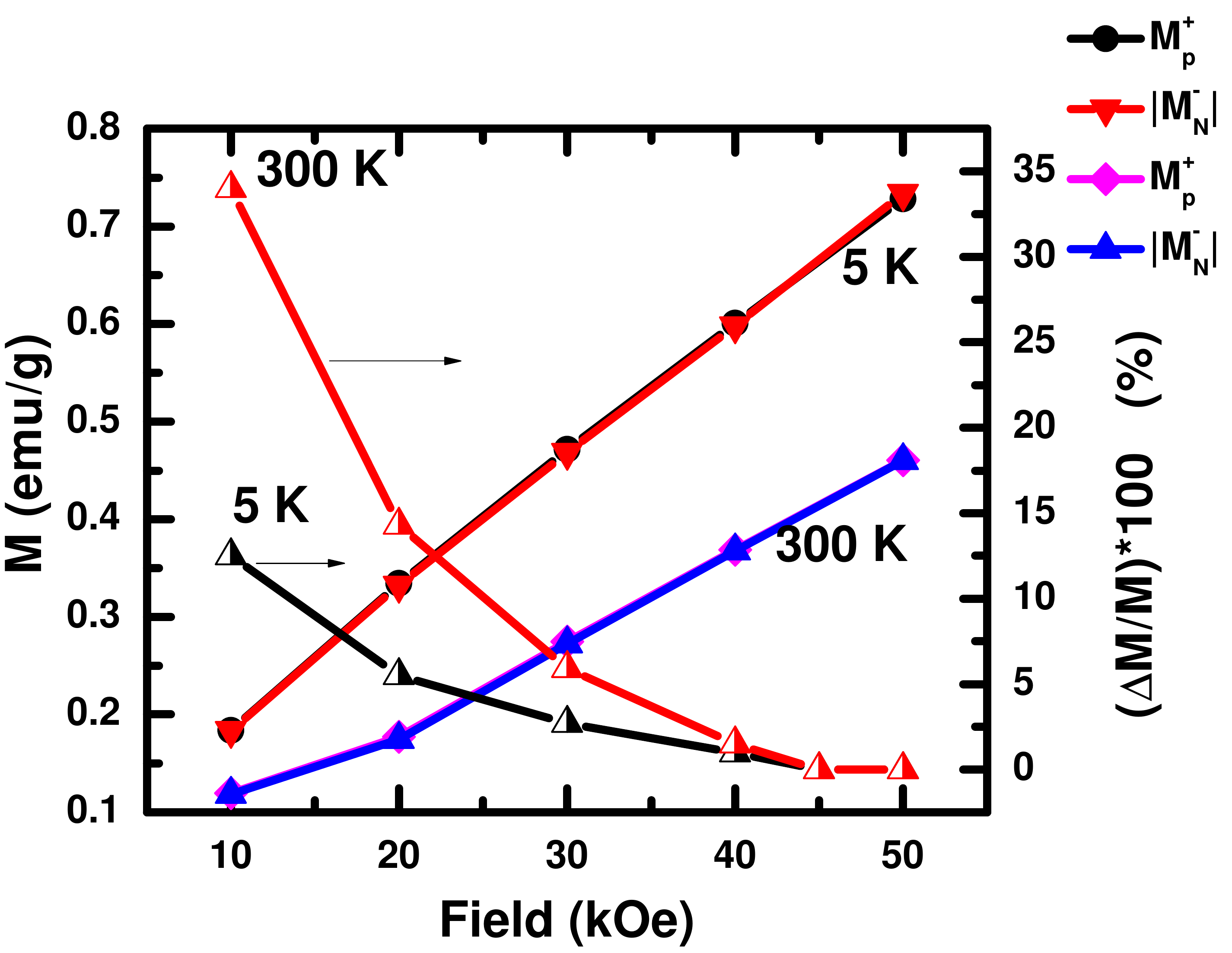} 
    \end{center}
  \caption{(color online) (a) The magnetization values at the positive and negative maximum fields ($H_m$) for different $H_m$s; the extent of irreversiblity as a function of magnetic fields for the loops measured at 5 and 300 K are also shown. }
\end{figure}

\begin{figure}[!h]
 \begin{center}
    \includegraphics[scale=0.30]{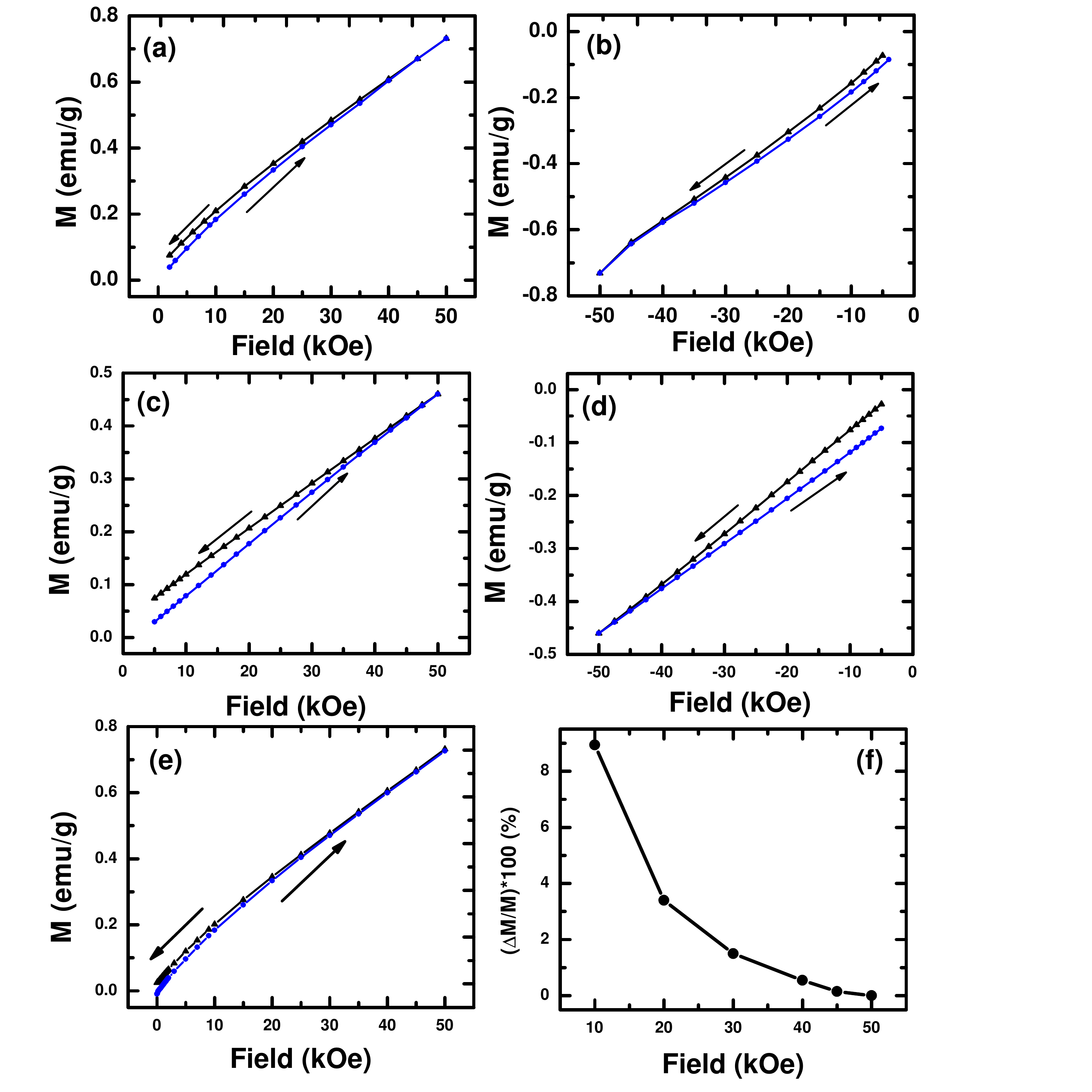} 
    \end{center}
  \caption{(color online) Certain portions of the hysteresis loops have been blown up to show how the forward and reverse branches of the loops merge around the maximum field (50 kOe) used in this work at different temperatures such as at (a), (b) 5 K and (c), (d) 300 K; the initial magnetization curve together with the reverse branch of the loop measured at 5 K are shown in (e); (f) shows the variation in the extent irreversibility with field between the initial and reverse branches measured at 5 K. Clearly, in all theses cases the magnetization becomes reversible around an applied field 40-45 kOe. These loops are not minor. }
\end{figure}

In response to the second point, we mention the following. The measurement of path dependency of the spontaneous exchange bias ($H_E$) has been carried out on the same sample. The sample has been appropriately demagnetized using oscillating magnetic field with decreasing amplitude to ensure identical initial magnetization state prior to each measurement. In fact, this issue has been rechecked by carrying out the measurements following zero field cooling from above 700 K (Supplementary document \cite{Maity-1}; see also Fig. 3 of this reply) which is above the magnetic transition points. Therefore, the initial magnetization states of the sample did not differ in the measurements carried out to examine the dependence of $H_E$ on the direction of the applied field. Since the path dependency has been clearly observed in the case of spontaneous exchange bias, its observation in the case of conventional exchange bias is indisputable.

\begin{figure}[!h]
  \begin{center}
    \includegraphics[scale=0.15]{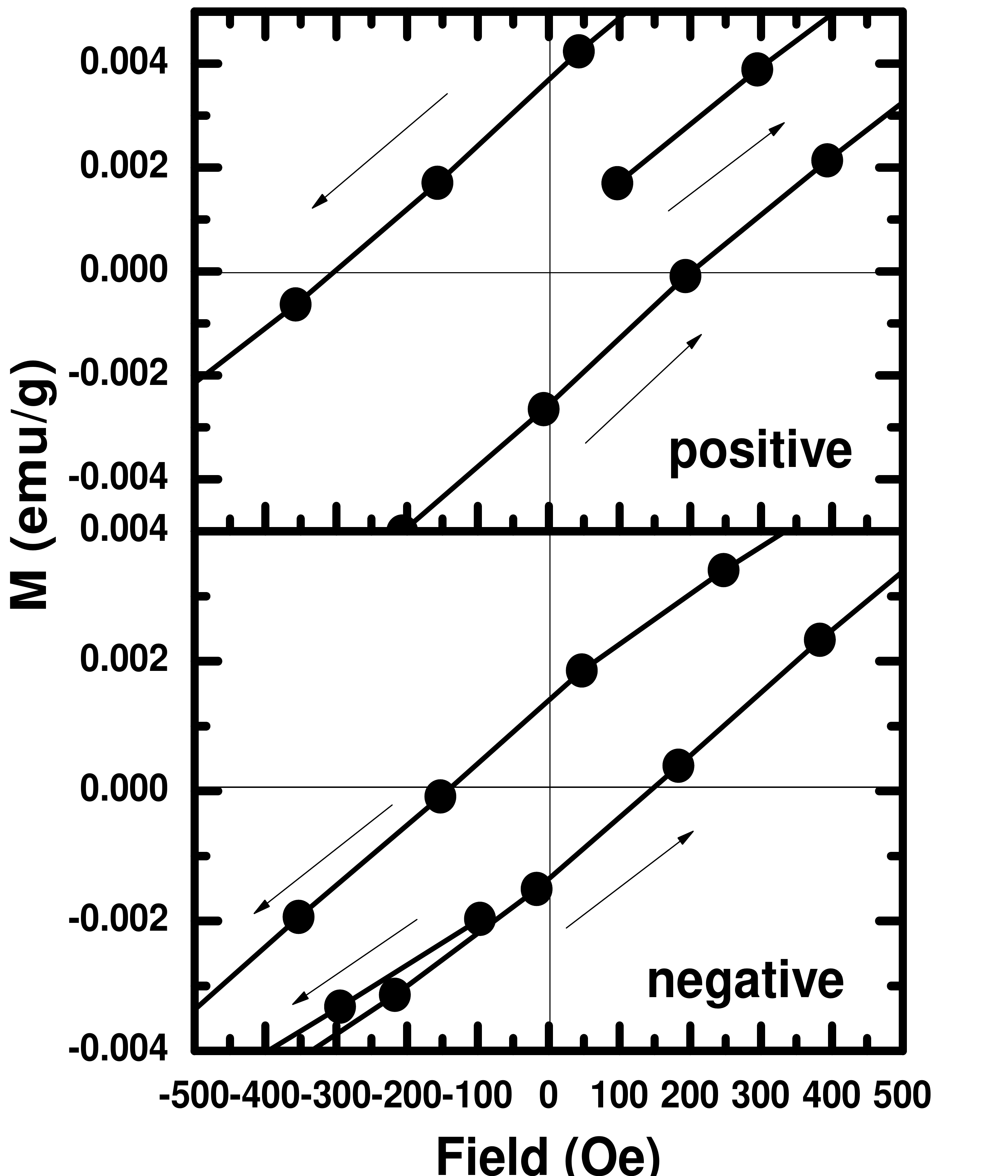} 
    \end{center}
  \caption{The blown-up portions of the hysteresis loops measured by vibrating sample magnetometer at room temperature following zero-field cooling from 700 K which is above the magnetic transition point of the system (~590 K). }
\end{figure}

The third point concerns about the presence of superspin glass phase. Although the memory effect has not been measured using the original protocol \cite{Sasaki}, the implication of which will be addressed separately, we point out that (i) we checked the fitting of the training curve by measuring the hysteresis loops with even smaller field step and found the validity of the equation (1); (ii) clear signature of spin glass transition (frequency dependence of the transition temperature in ac susceptibility measurement) could be noticed in a similar system \cite{Maity-2}; and, finally and more importantly, (iii) subtraction of the contribution of paramagnetic component from the overall field-cooled magnetization yields a temperature independent pattern \cite{Sasaki} at low temperature (Fig. 3c \cite{Maity-1}) offering $\textit{unambiguous}$ evidence for the presence of superspin glass phase in sample A. It is worth mentioining here that only sample A with higher volume fraction of superspin glass phase exhibits large spontaneous exchange bias.   

We further mention that a stronger ferromagnetic component is expected in even finer particles used in our study ($\sim$19 nm) than what has been used in the earlier work on nanoscale Bi$_2$Fe$_4$O$_9$ \cite{Tian,Zhang}. Finally, we point out that the error bar for the parameters such as H$_E$, H$_C$ etc is $\pm$1-3\%.

\end{document}